# Magnetic dipole absorption of light at intersubband transitions in quantum wells

I. V. Oladyshkin, M. A. Erukhimova, and M. D. Tokman

Institute of Applied Physics of the Russian Academy of Sciences, 603950, Nizhny Novgorod, Russia

We consider theoretically magnetic dipole mechanism of the IR-light absorption at intersubband transitions in wide-gap quantum wells (QW). In contrast to the electric dipole resonance, discussed mechanism manifests in the interaction with the *s*-polarized component of electromagnetic radiation. We demonstrated that a) the frequencies of magnetic and electric dipole intersubband transitions in 2D layer are equal in the framework of one-particle approximation; b) collective plasma effects should lead to a significant frequency shift between the magnetic and electric dipole resonances in typical QW. That is why independent measurements of the magnetic and electric dipole resonant frequencies could allow both the extracting the QW parameters and the experimental verifying of theoretical models of collective fermion interaction. Taking into account relative weakness of the magnetic dipole absorption, we propose a waveguide scheme for its experimental observation.

**keywords:** *quantum well, magnetic dipole resonance, intersubband transitions, light absorption*

**INTRODUCTION**

Physical systems with the size quantization (quantum wells, quantum wires and quantum dots) are currently under active investigation both due to the potential applications in nonlinear optics [1-3] and the variety of interesting fundamental effects. The latter phenomena are the bosonization of intersubband fermion oscillations [4, 5], the strong Stark splitting [6], the Purcell enhancement of spontaneous or induced quantum-optical processes [7, 8] and others.

In this paper we predict theoretically the effect of the resonant excitation of the orbital high-frequency magnetic moment in the plane of a quantum well (QW). In particular, the magnetic dipole interaction should lead to the absorption of the *s*-polarized (or normally incident) radiation. Note that the absorption of the *s*-polarized light can be also caused by multiband coupling effects in QW with a strong coupling between the conduction band and the valence band, i.e., in the narrow-gap structures [9-13]. The interaction mechanism considered in the present paper is completely different. The intensity of the magnetic dipole energy exchange between the field and the medium is determined by the ratio of the QW thickness to the characteristic spatial scale of the high-frequency field. Absorption rate is proportional to the squared amplitude of the high-frequency magnetic field projected onto the QW plane.

The listed features are quite typical for the magnetic dipole resonances. In this connection, we refer to the paper [14], where the magnetic dipole resonance was considered for a quantum dot. In the case of a spherically symmetric system, the magnetic and electric dipole resonances have different frequencies due to the corresponding selection rules (see [14] for details). In contrast, a plane 2D quantum layer, considered in the present paper, appeared to be a degenerate system since the frequencies of the both resonances are equal in the framework of one-particle approximation. However, the collective plasma effects remove this frequency degeneracy.

We showed that the plasma frequency shift of the electric dipole resonance may significantly exceed the homogeneous absorption broadening for a typical QW (see Section IV). Moreover, Coulomb interaction may cause the hybridization of different resonant transitions [4, 5, 28] and so-called "multisubband" plasmon excitation which leads to even stronger shift of the electric dipole resonances [28-30]. At the same time, we demonstrated that the corresponding shift of the magnetic dipole resonance is much less than the homogeneous broadening. That is why independent measurements of the magnetic dipole and electric dipole resonant frequencies could allow both the extracting the QW parameters and experimental verifying of theoretical models of collective fermion interaction. This result is promising from the experimental point of view despite the relative weakness of the magnetic dipole interaction.

Another nontrivial feature of the magnetic dipole resonance in QW is that in the case of degenerate electron distribution the absorption rate of light is proportional to the difference of the *squared* populations at two resonant levels (but not to the population difference as in the case of "common" electric dipole absorption).

In **Section I** the basic mechanism of the magnetic dipole interaction in QW is presented. In **Section II** we formulate the self-consistent model of the electromagnetic field interaction with free charge carriers in a thin layer. The carrier dynamics is described by the simplest QW model assuming rather large energy gap between the conductive band and the valence band. In **Section III** the high-frequency magnetic susceptibility of QW is derived in the rotating wave approximation (RWA) and the key properties of the magnetic and electric dipole resonances are compered basing on the self-consisted model. **Section IV** is devoted to the estimations and the comparison of the magnetic dipole absorption of IR radiation with other mechanisms existing in the case of an isolated conductive band.

## I. The basic model of magnetic dipole resonance in QW

Let us consider a quantum well lying in the plane ($x$, $y$), so the size quantization is in the "$z$" direction (see Fig. 1). Suppose that the conduction band and the valence band are separated by a rather large energy gap. In this case, the following eigenvectors can be attributed to the conduction electrons:

$$|m, \boldsymbol{k}> \equiv \Phi_m(z) e^{ik_x x + ik_y y}. \qquad (1)$$

Here $\Phi_m(z)$ corresponds to the subbands resulting from the size quantization. Suppose that during the time interval $t \in [0, T]$ an electron is under the action of electromagnetic field given by the vector potential $\boldsymbol{A} = \text{Re}\, \boldsymbol{x}_0 \tilde{A}_x(z) e^{-i\omega t}$ and the frequency $\omega$ is equal to the frequency of the intersubband transition between the states $\Phi_1(z) e^{ik_x x}$ and $\Phi_2(z) e^{ik_x x}$. Solving the Schrödinger equation by the perturbation method, for the initial state $\Phi_1(z) e^{ik_x x}$ at $t > T$ we obtain

$$\Psi(t, T) \approx \left[\Phi_1(z) e^{-i\frac{E_{1,k_x}}{\hbar}t} + C_2 \Phi_2(z) e^{-i\frac{E_{2,k_x}}{\hbar}t}\right] e^{ik_x x} + o(\tilde{A}_x^2), \qquad (2)$$

where $E_{2,k_x}$ and $E_{1,k_x}$ are the energies of "2" and "1" states correspondingly, $\omega = \frac{E_{2,k_x} - E_{1,k_x}}{\hbar}$, $C_2 = -i\frac{ek_x T}{2m_\| c}\tilde{A}_{21}$, $\tilde{A}_{21} = \langle \Phi_2|\tilde{A}_x|\Phi_1 \rangle$, $-e$ is the electron charge, $m_\|$ is the effective mass of the charge carrier in the plane of QW. The obtained state function $\Psi(t, T)$ gives the following current density in the "$x$" direction:

$$j_x \approx -\frac{e\hbar k_x}{m_\|}|\Phi_1(z)|^2 + \frac{e^2 \hbar k_x^2 T}{m_\|^2 c}\text{Re}\left[i\Phi_1^*(z)\Phi_2(z)\tilde{A}_{21}e^{-i\omega t}\right] + o(\tilde{A}_x^2). \qquad (3)$$

High-frequency component of the current depends on the quasi-momentum $k_x$ squared and, therefore, it doesn't disappear during the averaging over the equilibrium ensemble. If the functions $\Phi_1(z) e^{ik_x x}$ and $\Phi_2(z) e^{ik_x x}$ are of the different parity, then the high-frequency current have a stratified two-stream structure with the magnetic moment $\boldsymbol{m}$ oriented along the "y" axis (see Fig. 1).

It is easy to see that this effect is possible only if the vector potential $\tilde{A}_x$ is inhomogeneous along the $z$ axis, when the matrix element $\tilde{A}_{21} \neq 0$. For the relatively thin layer we get $\tilde{A}_x \approx \tilde{A}_x(0) + \left(\frac{\partial \tilde{A}_x}{\partial z}\right)_{z=0} z$, so the corresponding matrix element is $\tilde{A}_{21} = \tilde{H}_y z_{21}$, where $\tilde{H}_y$ is the complex amplitude of the magnetic field *y*-component. Note that in one-particle approximation the magnetic moment can be excited at the frequencies of intersubband transitions, which are the same as the frequencies of vertical polarization oscillations (along the *z* axis). From this point of view considered 2D system is degenerate, and this type of degeneracy is absent in 3D systems [14].

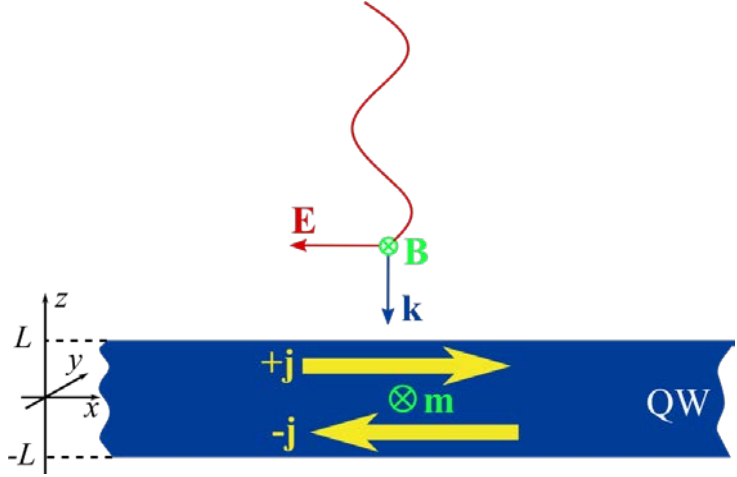

Fig. 1. Schematic picture of magnetic dipole absorption of *s*-polarized light. The simplest structure of electric currents inside the QW is shown by yellow arrows.

## II. Self-consistent model of QW in high-frequency field

### II.1 Self-consistent equations of the electromagnetic field in a thin layer

Let us consider a thin layer located in the area $-L \leq z \leq L$ between two half-spaces with the dielectric constants $\varepsilon_{(-)}$ for $z < -L$ and $\varepsilon_{(+)}$ for $z > L$. The permittivity of the layer determined by the valence electrons is $\varepsilon(z)$. Suppose that this QW contains free electrons with the density $n(\mathbf{r})$. The value of $n(\mathbf{r})$ is connected with the free charge density $\rho(\mathbf{r})$ by the relation $\rho(\mathbf{r}) = Q_i(z) - en(\mathbf{r})$, where $Q_i(z)$ is the ion charge density not compensated by the valence electrons.

We consider a QW occupying relatively small part of the "external" electrodynamic system. This can be simply a plane wave incident on a QW or a planar waveguide with a QW inside it (see Figs. 2 and 3). Thus, we set all the physical values inside the QW as $\propto e^{i\mathbf{qr}-i\omega t}$, where the vectors $\mathbf{q}$ and $\mathbf{r}$ belong to (*x,y*) plane and the frequency $\omega$ and the wavevector $\mathbf{q}$ are determined by the "external" eletrodynamical system. Dependencies of the considered physical values on the "*z*" coordinate are determined in a self-consistent way.

In Fig. 2 the incidence of electromagnetic waves of different polarizations on QW is shown. Here the vector $\mathbf{q}$ equals to the longitudinal projection of the incident radiation wavevector. In Fig. 3 a QW put in a waveguide is shown (the example of TE mode is given, which corresponds to the *s*-polarized field). Here the vector $\mathbf{q}$ is determined by the structure of a waveguide mode at given frequency $\omega$. All the calculations below are valid both for the cases of the incident plane wave and the waveguide propagation.

Keeping in mind further comparison of different interaction mechanisms, we consider both the *s*- and *p*-polarized light. Here it is more convenient to set the electromagnetic fields by the vector and scalar potentials of the following form: $\mathbf{A} = \mathrm{Re}\,\mathbf{x}_0 \tilde{A}_x(z) e^{i\mathbf{qr}-i\omega t}$ and $\varphi = \mathrm{Re}\,\tilde{\varphi}(z)e^{i\mathbf{qr}-i\omega t}$. In particular, for the *s*-polarized field we have $\varphi = 0$ and $\mathbf{q} \parallel \mathbf{y}_0$, and for the *p*-polarized field $\varphi \neq 0$ and $\mathbf{q} \parallel \mathbf{x}_0$ (here $\mathbf{x}_0$ and $\mathbf{y}_0$ are the unit vectors directed along the *x* and *y* axis respectively). Using this representation we got $A_{z,y} = 0$ for both polarizations, which significantly simplify the solution of the masters equations for the charge carriers in QW.

In the case of the *s*-polarized electromagnetic field (see Fig. 2, a and Fig. 3) it is enough to use the standard equation for the vector potential:

$$-\left(\frac{\partial^2}{\partial z^2} + \frac{\partial^2}{\partial y^2}\right)\tilde{A}_x(z) = \left(q^2 - \frac{\partial^2}{\partial z^2}\right)\tilde{A}_x(z) = \frac{\omega^2 \varepsilon(z)}{c^2}\tilde{A}_x(z) + \frac{4\pi}{c}\tilde{j}_x(z) = 0, \quad (4)$$

where $\tilde{j}_x(z)$ is the complex amplitude of the current $j_x = \mathrm{Re}\,\tilde{j}_x(z)e^{iqy-i\omega t}$.

Below we discuss quite common case of a thin QW satisfying inequalities:

$$qL, L\sqrt{\bar{\varepsilon}}\omega/c \ll 1 \quad (5)$$

where $\bar{\varepsilon}$ is the average value of $\varepsilon(z)$ into the layer. After the double integration on *z*, taking into account inequalities (5), we find:

$$\tilde{A}_x(z) - \tilde{A}_x(-L) = -\frac{4\pi}{c}\int_{-L}^{z} dz' \int_{-L}^{z'} dz'' \tilde{j}_x(z'') + \tilde{B}_y(-L) \cdot (z+L). \tag{6}$$

For the *p*-polarized field perturbations of all values, including the spatial charge perturbation, can be set as $f = \text{Re}\,\tilde{f}(z)e^{iqx-i\omega t}$ (see Fig. 2b). For such field the self-consistent system of equations should include:

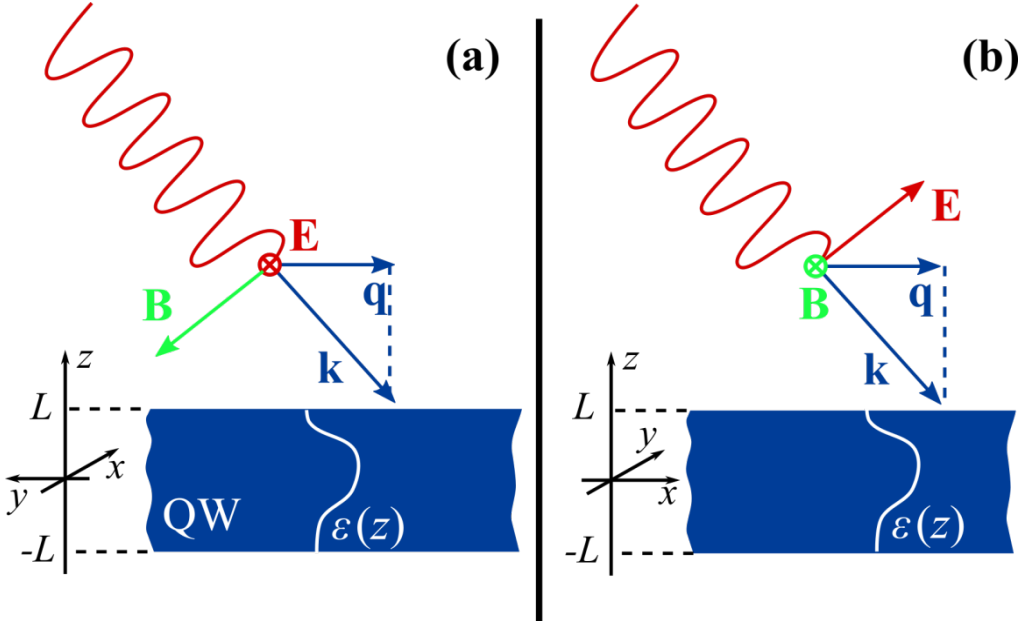

Fig. 2. Incidence of (a) s-polarized and (b) p-polarized waves at QW.

(**i**) the Gauss's law $\nabla \boldsymbol{D} = 4\pi\rho$:
$$\frac{\partial}{\partial z}\left(\varepsilon(z)\frac{\partial \tilde{\varphi}}{\partial z}\right) = \varepsilon(z)q^2\tilde{\varphi} - \varepsilon(z)q\frac{\omega}{c}\tilde{A}_x - 4\pi\tilde{\rho}, \tag{7}$$
where $\boldsymbol{D} = \varepsilon(z)\left(-\nabla\varphi - \frac{1}{c}\dot{\boldsymbol{A}}\right)$ is the electric displacement field. Note that the spatial charge perturbation $\tilde{\rho}$ in Eq. (7) is the perturbation relative to the stationary background, which forms the QW potential profile.

(**ii**) the continuity equation $\dot{\rho} + \nabla \boldsymbol{j} = 0$:
$$i\omega\tilde{\rho} = \frac{\partial}{\partial z}\tilde{j}_z + iq\tilde{j}_x, \tag{8}$$
(**iii**) the *z*-component of the Maxwell equation $\nabla \times \boldsymbol{B} = 4\pi\boldsymbol{j}/c + \dot{\boldsymbol{D}}/c$:
$$q\frac{\partial \tilde{A}_x}{\partial z} = -\frac{4\pi i}{c}\tilde{j}_z + \frac{\omega\varepsilon(z)}{c}\frac{\partial \tilde{\varphi}}{\partial z}. \tag{9}$$
From Eq. (7), taking into account inequalities (5), we obtain the following equation for the complex amplitude of the scalar potential oscillation:
$$\tilde{\varphi}(z) = -\tilde{D}_z(-L)\int_{-L}^{z}\frac{dz'}{\varepsilon(z')} - 4\pi\int_{-L}^{z}\frac{dz'}{\varepsilon(z')}\int_{-L}^{z'}\tilde{\rho}(z'')dz''. \tag{10}$$
After that we can substitute Eqs. (8), (10) in Eq. (9) and integrate it. Using boundary conditions $j_z(\pm L) = 0$ and $\tilde{D}_z(\pm L) = -\frac{qc}{\omega}\tilde{B}_y(\pm L)$, we obtain equation for the vector potential in the same form as Eq.(6).

## II.2 Dynamics of free charge carriers

Let us define the potential well for electrons as $U_{qw}(z)$ and use the simplest model for the electron Hamiltonian assuming that the energy gap between the valence and conduction bands is enough large [1,15,16]:

$$\hat{H} = -eU_{qw}(z) + \frac{\hat{p}_x^2}{2m_\parallel} + \frac{\hat{p}_y^2}{2m_\parallel} + \frac{\hat{p}_z^2}{2m_\perp} + \hat{V}, \tag{11}$$

where $\hat{\boldsymbol{p}} = -i\hbar\nabla$ is the momentum operator, $m_{\perp,\parallel}$ are the components of the effective mass tensor, $\hat{V}$ is the interaction operator. For the chosen gauge ($A_{z,y} = 0$) operator $\hat{V}$ has the form:

$$\hat{V} = -e\varphi + e\frac{\hat{p}_x A_x + A_x \hat{p}_x}{2m_\parallel c} + \frac{e^2 A_x^2}{2m_\parallel c^2}, \tag{12}$$

where $\hat{\boldsymbol{p}} = -i\hbar\nabla$ is the momentum operator, $m_{\perp,\parallel}$ are the components of the effective mass tensor, $\hat{V}$ is the interaction operator (recall that for the *s*-polarized field $\varphi = 0$).

In the absence of the perturbative potentials $\varphi$ and $\boldsymbol{A}$ the Hamiltonian (11) describes formation of subbands with the energies $E_{m,\boldsymbol{k}} = E_m + \frac{k^2\hbar^2}{2m_\parallel}$ and the following eigenfunctions:

$$|m,\boldsymbol{k}\rangle = \Psi_{m\boldsymbol{k}}(\boldsymbol{r}) = \Phi_m(z)\exp(ik_x x + ik_y y), \tag{13}$$

where the functions $\Phi_m(z)$ and values $E_m$ are determined by the one-dimensional Schrödinger equation:

$$E_m\Phi_m = -\frac{\hbar^2}{2m_\perp}\frac{\partial^2\Phi_m}{\partial z^2} - eU_{qw}(z)\Phi_m, \tag{14}$$

while the set of quasi-momenta $\boldsymbol{k} = (k_x, k_y)$ is determined by the periodic boundary conditions at the border of a unit area.

Under the condition $(\bar{\varepsilon}\omega/c)/k_F \ll 1$ ($k_F$ is the Fermi quasi-momentum) the problem can be reduced to the study of "spinless" carriers, while the spin degeneration can be taken into account just in the macroscopic values calculations (see **Appendix** for details).

The average carrier density $n(\boldsymbol{r})$ and the current $\boldsymbol{j}(\boldsymbol{r})$ are determined by (e.g., see [17, 18]):

$$n(\boldsymbol{r}) = \sum_{\alpha\beta}\Psi_\beta^*(\boldsymbol{r})\Psi_\alpha(\boldsymbol{r})\rho_{\alpha\beta}, \boldsymbol{j}(\boldsymbol{r}) = -\frac{e}{2}\sum_{\alpha\beta}\{\Psi_\beta^*(\boldsymbol{r})[\hat{v}\Psi_\alpha(\boldsymbol{r})] + [\hat{v}^*\Psi_\beta^*(\boldsymbol{r})]\Psi_\alpha(\boldsymbol{r})\}(\boldsymbol{r})\rho_{\alpha\beta}, \tag{15}$$

where $\rho_{\alpha\beta}$ is the density matrix and $|\alpha\rangle \equiv |m,\boldsymbol{k}\rangle$; $\hat{v} = \frac{i}{\hbar}[\hat{H},\boldsymbol{r}]$ is the velocity operator[1] [15].

The density matrix is described by the von Neumann equation:

$$\frac{\partial\rho_{\alpha\beta}}{\partial t} + i\frac{E_\alpha - E_\beta}{\hbar}\rho_{\alpha\beta} + \frac{i}{\hbar}\sum_\chi(V_{\alpha\chi}\rho_{\chi\beta} - \rho_{\alpha\chi}V_{\chi\beta}) = 0. \tag{16}$$

Solving Eq. (16) we will use the simplest model of relaxation processes using the substitution $\omega \Rightarrow \omega + i\gamma$ in the resonant terms $\propto \frac{1}{E_\alpha - E_\beta - \hbar\omega}$ [19, 20]. If the relaxation constant $\gamma$ is much less than the resonant frequency $\omega_{\alpha\beta} = \frac{|E_\alpha - E_\beta|}{\hbar}$, this approach is valid for the description of the resonant effects within the framework of RWA approximation applicability – see papers [21-23] for details.

From the first equality in Eq. (15) we obtain the oscillating component of the electron density, using the matrix elements of the density operator perturbation relative to the equilibrium value. The complex amplitude of these oscillations $\tilde{n}$ is directly connected with the previously introduced in Eq. (7) complex amplitude of the spatial charge oscillations as $\tilde{\rho} = -e\tilde{n}$.

## III. Electric dipole and magnetic dipole oscillations in QW

Let us consider a linear electromagnetic response of QW and neglect quadratic terms $\sim A_x^2$ in the interaction operator $\hat{V}$ (see Eq. (12)). For the harmonic dependence $\propto e^{\pm i\boldsymbol{q}\boldsymbol{r}}$ (the vectors $\boldsymbol{q}$ and $\boldsymbol{r}$ belong to *x,y* plane) the matrix elements of the interaction operator have the form $V_{mnkk'} \propto \delta_{(\boldsymbol{k}\mp\boldsymbol{q})\boldsymbol{k}'}$. Under the condition $q/k_F \ll 1$ the interband transitions are "direct" and, consequently, $V_{mnkk'} \approx \delta_{kk'}V_{mnkk}$, which can be presented as

$$V_{mnkk'} = \delta_{kk'}\frac{1}{2}(\tilde{V}_{mnkk}e^{-i\omega t} + H.C), \tag{17}$$

---

[1] It is easy to see that Eq.(15) corresponds to the standard expression for the matrix elements of the current density operator: $\boldsymbol{j}_{\alpha\beta} = -\frac{e}{2m_\parallel}[\Psi_\alpha^*(\hat{\boldsymbol{p}}\Psi_\beta) + (\hat{\boldsymbol{p}}^*\Psi_\alpha^*)\Psi_\beta] - \frac{e^2\boldsymbol{A}}{m_\parallel c}\Psi_\alpha^*\Psi_\beta$ (see [15]).

where $\tilde{V}_{mnkk} = -e\tilde{\varphi}_{mn} + \frac{e\hbar k_x}{m_\parallel c}\tilde{A}_{mn}$, $f_{mn} = \langle \Phi_m | f | \Phi_n \rangle$. Since the interaction operator $V_{mnkk'}$ assumed to be diagonal with respect to $k\, k'$ indices, the density matrix perturbation can be represented in a similar way:

$$\rho_{mnkk'} = \delta_{mn}\delta_{kk'}n_{mk}^{(0)} + \delta_{kk'}\frac{1}{2}\left(\tilde{\rho}_{mnkk}e^{-i\omega t} + H.C\right), \tag{18}$$

where $n_{mk}^{(0)}$ are equilibrium (unperturbed) populations. From Eqs. (16) and (17),(18) we find the linear perturbation of the density matrix:

$$\tilde{\rho}_{mnkk} = \frac{n_{nk}^{(0)} - n_{mk}^{(0)}}{\omega_{mn} - \omega - i\gamma}\left(\frac{e\tilde{\varphi}_{mn}}{\hbar} - \frac{ek_x\tilde{A}_{mn}}{m_\parallel c}\right). \tag{19}$$

Eqs. (15) and (19) allow to obtain complex amplitudes of the longitudinal current and electron density. Still using RWA approximation, we will take into account just one intersubband transition with the frequency $\omega_{mn} \approx \omega$. For distributions $n_{(m,n)k}^{(0)}$ symmetric with respect to the longitudinal quasi-momentum $k_x$ we come to the following expressions:

$$\tilde{j}_x(z) = \frac{e^2\hbar}{m_\parallel^2 c}\sum_k \frac{k_x^2\left(n_{nk}^{(0)} - n_{mk}^{(0)}\right)}{\omega_{mn} - \omega - i\gamma}\Phi_n^*(z)\Phi_m(z)\tilde{A}_{mn}, \tag{20}$$

$$\tilde{n}(z) = \frac{e}{\hbar}\sum_k \frac{\left(n_{nk}^{(0)} - n_{mk}^{(0)}\right)}{\omega_{mn} - \omega - i\gamma}\Phi_n^*(z)\Phi_m(z)\tilde{\varphi}_{mn}. \tag{21}$$

To find the matrix elements $\tilde{\varphi}_{mn}$ and $\tilde{A}_{mn}$ in a self-consistent way we use Eqs. (6), (10) and find:

$$\tilde{\varphi}_{mn} = \tilde{D}_z(-L)\frac{d_{mn}^{(eff)}}{e} + 4\pi e\left\langle \Phi_m \left| \int_{-L}^{z}\frac{dz'}{\varepsilon(z')}\int_{-L}^{z'}\tilde{n}(z'')dz'' \right| \Phi_n \right\rangle, \tag{22}$$

$$\tilde{A}_{mn} - \delta_{mn}\tilde{A}_x(-L) = -\frac{4\pi}{c}\left\langle \Phi_m \left| \int_{-L}^{z}dz'\int_{-L}^{z'}dz''\,\tilde{j}_x(z'') \right| \Phi_n \right\rangle + $$
$$+ \tilde{B}_y(-L)\cdot(z_{mn} + \delta_{mn}L); \tag{23}$$

here $d_{mn}^{(eff)} = -e\left\langle \Phi_m \left| \int_0^z \frac{dz'}{\varepsilon(z')} \right| \Phi_n \right\rangle$. Substitution of Eqs. (20), (21) in Eqs. (22), (23) gives the self-consistent solutions for the matrix elements $\tilde{\varphi}_{mn}$ and $\tilde{A}_{mn}$ which includes the fields at the QW boundary $\tilde{D}_z(-L)$ and $\tilde{B}_y(-L)$.

Complex amplitudes $\tilde{j}_x(z) \propto \tilde{n}(z) \propto \Phi_n^*\Phi_m$ meet the condition $\int_{-L}^{L}\tilde{j}_x(z)\,dz = \int_{-L}^{L}\tilde{n}(z)\,dz = 0$, which leads to $\tilde{B}_y(-L) = \tilde{B}_y(+L) = \tilde{H}_y$ and $\tilde{D}_z(-L) = \tilde{D}_z(+L) = \tilde{D}_z$:

$$\tilde{\varphi}_{mn} = \frac{\omega_{mn} - \omega - i\gamma}{\omega_{mn} + \Omega_{mn}^{(\varphi)} - \omega - i\gamma}\cdot\frac{d_{mn}^{(eff)}}{e}\tilde{D}_z, \tag{24}$$

$$\tilde{A}_{mn} = \frac{\omega_{mn} - \omega - i\gamma}{\omega_{mn} + \Omega_{mn}^{(A)} - \omega - i\gamma}\cdot z_{mn}\tilde{H}_y, \tag{25}$$

where

$$\Omega_{mn}^{(\varphi)} = \frac{4\pi e^2(N_n - N_m)G_{mnnm}}{\hbar} \quad \text{and} \quad \Omega_{mn}^{(A)} = -\frac{4\pi e^2(N_n - N_m)J_{mnnm}}{\hbar}\frac{W_{eff}^{mn}}{m_\parallel c^2} \tag{26}$$

are the frequency shifts due to collective effects;

$$J_{mnnm} = \frac{\hbar^2}{4\omega_{mn}^2 m_\perp^2}\int_{-L}^{L}\left|\Phi_m(z)\frac{\partial \Phi_n^*}{\partial z} - \Phi_n^*(z)\frac{\partial \Phi_m(z)}{\partial z}\right|^2 dz, \tag{27}$$

$$G_{mnnm} = \frac{\hbar^2}{4\omega_{mn}^2 m_\perp^2}\int_{-L}^{L}\frac{1}{\varepsilon(z)}\left|\Phi_m(z)\frac{\partial \Phi_n^*}{\partial z} - \Phi_n^*(z)\frac{\partial \Phi_m(z)}{\partial z}\right|^2 dz \tag{28}$$

are so-called overlap integrals[2];

$$W_{eff}^{mn} = 2\frac{N_n\langle W_{\|n}\rangle - N_m\langle W_{\|m}\rangle}{N_n - N_m}, \quad N_n = \sum_{\mathbf{k}} n_{n\mathbf{k}}^{(0)}, \quad \langle W_{\|n}\rangle = \frac{1}{N_n}\sum_{\mathbf{k}} \frac{\hbar^2 k_x^2}{2m_\|} n_{n\mathbf{k}}^{(0)}. \tag{29}$$

Note that for the *s*-polarized field we have $\tilde{\varphi}_{mn} = \widetilde{D}_z = 0$. The last of Eqs. (29) determines the average energies of the longitudinal motion (for the band under consideration).

In Eqs. (29) the summation can be replaced by the integration: $\sum_{\mathbf{k}}(\ldots) \Rightarrow \frac{g}{4\pi^2}\iint_\infty (\ldots) d^2k$, where $g = 2$ is the spin degeneracy factor. For the degenerate distribution with the Fermi energy $E_F$ we obtain:

$$N_n = \frac{m_\|(E_F - E_n)}{\pi\hbar^2}, \quad \langle W_{\|n}\rangle = \frac{\pi\hbar^2 N_n}{4m_\|}, \quad W_{eff}^{mn} = \frac{\pi\hbar^2}{m_\|}\frac{N_n + N_m}{2}, \tag{30}$$

for $E_m > E_F > E_n$ we have $N_m = 0, N_n \neq 0$.

Note that in the absence of any external fields (when $\widetilde{D}_z = \widetilde{H}_y = 0$) Eqs. (24) and (25) determine the eigenfrequencies of the electric dipole and magnetic dipole oscillations respectively:

$$\tilde{\varphi}_{mn}\left(\omega_{mn} + \Omega_{mn}^{(\varphi)} - \omega - i\gamma\right) = 0, \tag{31}$$

$$\tilde{A}_{mn}\left(\omega_{mn} + \Omega_{mn}^{(A)} - \omega - i\gamma\right) = 0. \tag{32}$$

Using Eqns. (19), (24) and (25) we find the self-consistent solution for the density matrix:

$$\tilde{\rho}_{mn\mathbf{k}\mathbf{k}} = \frac{d_{mn}^{(eff)}\left(n_{n\mathbf{k}}^{(0)} - n_{m\mathbf{k}}^{(0)}\right)}{\hbar\left(\omega_{mn} + \Omega_{mn}^{(\varphi)} - \omega - i\gamma\omega\right)}\widetilde{D}_z + \frac{d_{mn}k_x\left(n_{n\mathbf{k}}^{(0)} - n_{m\mathbf{k}}^{(0)}\right)}{m_\| c\left(\omega_{mm} + \Omega_{mn}^{(A)} - \omega - i\gamma\right)}\widetilde{H}_y, \tag{33}$$

where $d_{mn} = -ez_{mn}$. From Eqs. (31)–(33) it follows that the electric dipole oscillations are excited exclusively by *p*-polarized radiation, containing the electric field component $\tilde{E}_z$, while the magnetic dipole oscillations are excited by the *s*-polarized radiation.

The first term in Eq. (33) describes the density matrix perturbation caused by an external electric field. This expression differs from the well-known solution for a two-level system [20]. First, there is a factor $d_{mn}^{(eff)}\widetilde{D}_z$ instead of $d_{nm}\tilde{E}_z$ which is due to the possible inhomogeneity of the dielectric permittivity into QW [7]. Second, there is a new term $\Omega_{mn}^{(\varphi)}$ in the denominator – so-called "depolarization shift" of the resonant frequency relative to the intersubband transition frequency [24].

The second term in Eq. (33) describes the excitation of the magnetic dipole oscillations in QW. And there is also a frequency shift $\Omega_{mn}^{(A)}$ caused by the collective effects.

Let us calculate the complex amplitudes of the resonant oscillations of the electric dipole and magnetic dipole moments per unit area of QW:

$$\tilde{P}_z = -e\int_{-L}^{L} z\tilde{n}dz, \quad \widetilde{M}_y = \int_{-L}^{L} \widetilde{m}_y dz, \tag{34}$$

where $-c\frac{\partial \widetilde{m}_y}{\partial z} = \tilde{j}_x$. Using Eq. (33) with Eqns. (20), (21), (24) and (25) we find:

$$\tilde{P}_z = \frac{d_{nm}d_{mn}^{(eff)}(N_n - N_m)}{\hbar\left(\omega_{mn} + \Omega_{mn}^{(\varphi)} - \omega - i\gamma\right)} \cdot \widetilde{D}_z, \tag{35}$$

$$\widetilde{M}_y = \frac{|d_{mn}|^2(N_n - N_m)}{\hbar\left(\omega_{mn} + \Omega_{mn}^{(A)} - \omega - i\gamma\right)} \cdot \frac{W_{eff}^{mn}}{m_\| c^2} \cdot \widetilde{H}_y. \tag{36}$$

Representing Eq. (36) as $\widetilde{M}_y = \chi^{(m)}\widetilde{H}_y$, we obtain the effective magnetic susceptibility at the resonant intersubband transition:

$$\chi^{(m)} = \frac{|d_{mn}|^2(N_n - N_m)}{\hbar\left(\omega_{mn} + \Omega_{mn}^{(A)} - \omega - i\gamma\right)} \cdot \frac{W_{eff}^{mn}}{m_\| c^2}. \tag{37}$$

---

[2] To obtain Eqs. (27), (28) we use the following property of stationary solutions of the Schrödinger equation: $\Phi_n^*\Phi_m = \frac{(\hbar/m_\perp)}{2\omega_{mn}}\frac{\partial}{\partial z}\left(\Phi_m\frac{\partial \Phi_n^*}{\partial z} - \Phi_n^*\frac{\partial \Phi_m}{\partial z}\right)$.

Thus, the resonant magnetic susceptibility $\chi^{(m)}$ differs from the electric susceptibility by the factor $\frac{W_{eff}^{mn}}{m_\| c^2}$ and by the value of resonant frequency (other differences are not so important). For a degenerate system the magnetic susceptibility (37) can be rewritten using the last of Eqs.(30):

$$\chi^{(m)} = \frac{\pi\hbar |d_{mn}|^2 (N_n^2 - N_m^2)}{2m_\|^2 c^2 \left(\omega_{mn} + \Omega_{mn}^{(A)} - \omega - i\gamma\right)}. \tag{38}$$

From Eq. (38) it follows that the magnetic susceptibility of QW is proportional to the difference of the squared carrier densities $(N_n^2 - N_m^2)$, but not to the difference of the densities themselves $(N_n - N_m)$ as the electric permittivity does. Also note that $\chi^{(m)}$ is grater for the charge carriers with the small "longitudinal" effective mass $m_\|$.

The power of magnetic dipole resonant losses per unit area of QW can be calculated using the initial Eq. (16) for the density matrix:

$$W = \frac{\partial}{\partial t}\sum_\alpha E_\alpha \rho_{\alpha\alpha} = \frac{2}{\hbar}\text{Im}\sum_{\alpha\beta} E_\alpha V_{\alpha\beta}\rho_{\beta\alpha}. \tag{39}$$

We still take into account just a resonant intersubband transition with $\omega_{mn} \approx \omega$; in this case substitution of Eqs. (17) and (18) in Eq. (39) gives:

$$W = \frac{\omega}{2}\text{Im}\sum_{\mathbf{k}} \tilde{V}_{mn\mathbf{k}\mathbf{k}}(\tilde{\rho}_{mn\mathbf{k}\mathbf{k}})^*. \tag{40}$$

Now we use in Eq. (40) previously derived Eqs. (17), (19) and (25) for the matrix elements of the interaction operator, for the density operator and for the vector potential respectively. We also put $\tilde{\varphi} = \tilde{D}_z = 0$ in the listed relations. The result confirms the well-known phenomenological formula [25] for the high-frequency field losses in magnetic medium:

$$W^{(m)} = \frac{\omega}{2}\text{Im}\chi^{(m)}|\tilde{H}_y|^2. \tag{41}$$

To conclude this section we should note that the discussed effect of the magnetic dipole absorption has a very close classical analogue: it is the electron-cyclotron absorption of an ordinary wave (O-mode), propagating perpendicularly to the external magnetic field in magnetized plasma. In this case the electric field of the wave is parallel to the constant magnetic field (and *does not rotate synchronously with electrons*). Nevertheless, the radiation of this polarization type can make nonzero contribution to the electrons' rotation energy under the cyclotron resonant conditions and the absorbed energy will be proportional to the squared ratio of gyroradius and wavelength [26, 27]. Thus, in this example the cyclotron oscillations of electrons play the same role as the intersubband transitions in QW in the case of the s-polarized radiation absorption.

## IV. Estimations

For the estimations we consider the wide-gap semiconductor GaAs, having $m_\perp \approx m_\| = m_{eff} = 0.067 m_0$ and $\bar\varepsilon \approx 10$. Let us choose the IR photon energy $\hbar\omega \approx 0.3$ eV (at which $\hbar\gamma \approx 10\ meV$) and consider the resonant transition between the two lowest levels, so $\omega = \omega_{21}$. For the transition between two lowest levels in a deep "rectangular" well this frequency corresponds to QW thickness $2L \approx 7 nm$ and $|z_{21}| \approx 1.2\ nm$.

The resonant frequency shifts $\Omega_{mn}^{(\varphi,A)}$ are related as $\Omega_{mn}^{(A)} \approx -\Omega_{mn}^{(\varphi)}\frac{\bar\varepsilon W_{eff}^{mn}}{m_\| c^2}$ and $\Omega_{mn}^{(\varphi)}$ can be expressed as

$$\frac{\Omega_{mn}^{(\varphi)}}{\omega_{mn}} \approx \alpha \frac{\lambda_{mn}}{L} \frac{2G_{mnnm}}{\pi L}\left(\frac{\omega_{mn} L^2 m_\perp}{\hbar}\right)\frac{m_\|}{m_\perp}\frac{N_n - N_m}{N^*} \tag{42}$$

where $\alpha = \frac{e^2}{\hbar c}$ is the fine structure constant, $\lambda_{mn} = 2\pi c/\omega_{mn}$ is the resonant radiation wavelength, $N^* = \frac{m_\| \omega_{mn}}{\pi\hbar}$ is the maximal electron density in the *n*-th (lower) subband under the condition that the *m*-th subband is empty.

The transition between two lowest levels in a deep "rectangular" well corresponds to $G_{mnnm} = \frac{5\pi^2}{32\bar\varepsilon}\frac{\hbar^2}{\omega_{mn}^2 m_\perp^2 L^3}$, which leads to the following form of Eq.(42):

$$\frac{\Omega_{mn}^{(\varphi)}}{\omega_{mn}} = \frac{5\pi}{16\bar{\varepsilon}}\alpha\frac{\lambda_{mn}}{L}\left(\frac{\hbar}{\omega_{mn}m_\perp L^2}\right)\frac{m_\parallel}{m_\perp}\frac{N_n - N_m}{N^*},$$

where $\frac{\hbar}{\omega_{mn}m_\perp L^2} \sim \frac{|z_{mn}|^2}{L^2}$. For the considered model of QW we obtain $\lambda_{21}/L \cong 1.15 \cdot 10^3$, $\frac{\hbar}{\omega_{21}m_\perp L^2} \approx 2\frac{|z_{21}|^2}{L^2} \approx 0{,}25$, which gives:

$$\frac{\Omega_{21}^{(\varphi)}}{\omega_{21}} \approx \frac{2.1}{\bar{\varepsilon}} \cdot \frac{m_\parallel}{m_\perp} \cdot \frac{N_1 - N_2}{N^*}.$$

One can see that the frequency shift of the electric dipole resonance can be significantly larger than the absorption line width even when the lower subband has low population.

Of course, at sufficiently high densities, the RWA approximation can be used only for rough estimations of electric dipole resonance characteristics. In the papers [4, 5, 28] the expression for the "intersubband" plasmon frequency was obtained: $\omega_{mn}^{(ip)} = \sqrt{\omega_{mn}^2 + \omega_{p_{mn}}^2}$, where $\omega_{p_{mn}}^2 = 2\omega_{mn}\Omega_{mn}^{(\varphi)}$ is the effective plasma frequency and $\Omega_{mn}^{(\varphi)}$ assumed to be comparable to $\omega_{mn}$. In the case of nonzero population at several subbands, the effect of the plasmon hybridization takes place. This leads to so-called "multisubband" plasmon excitation which cause even stronger shift of the electric dipole resonances [28-30].

At the same time, the resonant frequency shift of the magnetic dipole resonance $\left|\Omega_{mn}^{(A)}\right| \approx \Omega_{mn}^{(\varphi)}\frac{W_{eff}^{mn}}{\bar{\varepsilon}m_\parallel c^2}$ is much smaller than the intersubband transition frequency even at high carrier densities, when $\Omega_{mn}^{(\varphi)} \sim \omega_{mn}$. Moreover, for the values $\omega_{mn}/2\gamma \sim 10 \div 20$ [31] the shift is several orders of magnitude smaller than the resonance width.

To compare absolute values of electric and magnetic dipole absorption, we use Eq. (38) for the magnetic susceptibility and obtain the electric susceptibility from Eq. (35) for the homogeneous permittivity $\bar{\varepsilon}$ inside the layer:

$$\chi^{(e)} = \frac{|d_{mn}|^2(N_n - N_m)}{\hbar\left(\omega_{mn} + \Omega_{mn}^{(\varphi)} - \omega - i\gamma\right)}. \tag{43}$$

Thus, the ratio of the magnetic and electric dipole losses can be calculated from Eqs. (38) and (43). To make the simplest estimation we may consider a single propagation of the optical wave through QW, assuming $\varepsilon_{(-)} \approx \varepsilon_{(+)} \approx \bar{\varepsilon}$. In this case the ratio of the magnetic and electric dipole absorption at the corresponding resonant frequencies can be written as:

$$\frac{W^{(m)}}{W^{(e)}} \approx \frac{[\text{Im}\chi^{(m)}]_{\omega=\omega_{mn}+\Omega_{mn}^{(A)}}}{[\text{Im}\chi^{(e)}]_{\omega=\omega_{mn}+\Omega_{mn}^{(\varphi)}}}\frac{|\tilde{H}_y|^2}{|\tilde{E}_z|^2} \approx \frac{\pi\hbar^2(N_n + N_m)\varepsilon_{(-)}}{2m_\parallel^2 c^2}.$$

Assuming that $N_n + N_m \sim N^*$, we find: $\frac{W^{(m)}}{W^{(e)}} \sim \frac{\hbar\omega_{mn}\varepsilon_{(-)}}{m_\parallel c^2} \sim \frac{|z_{nm}|^2\omega^2}{c^2}\varepsilon_{(-)}$. For the listed parameters we obtain $\frac{\hbar\omega_{21}\varepsilon_{(-)}}{m_\parallel c^2} \sim 10^{-4}$, which corresponds, for example, to 0.1% absorption at the magnetic dipole resonance in 100-layer QW structure.

These estimations demonstrate that the magnetic dipole absorption of IR light is hard to measure in a single-pass geometry shown in Fig. 1. In this regard a waveguide scheme seems to be much more perspective, since it provides, in fact, multiple transmission of radiation through the layer (see Fig. 3).

Let us consider $TE_{N0}$ mode in the rectangular waveguide having the size $2l_\perp$ along the $z$ axis; $TE_{20}$ mode is shown in Fig. 3 as an example. In this case the electromagnetic field can be described by the following vector potential:

$$\boldsymbol{A} = \boldsymbol{x}_0\tilde{A}_x(z)e^{iqy-i\omega t} = \boldsymbol{x}_0\tilde{A}_0\begin{cases}\sin\left(\frac{N_{even}\pi}{2l_\perp}z\right)\\ \cos\left(\frac{N_{odd}\pi}{2l_\perp}z\right)\end{cases}e^{iqy-i\omega t}, \tag{44}$$

where $\kappa_\perp^2 + q^2 = \bar{\varepsilon}\frac{\omega^2}{c^2}$, $\kappa_\perp = \frac{N\pi}{2l_\perp}$. Let us place QW of the thickness $2L$ ($l_\perp \gg L$) along the wave knot plane $z = z_{knot}$, where $\mathbf{A} = \mathbf{E} = 0$. For simplicity, we neglect the influence of QW on the structure of the waveguide mode, although this is not a principal assumption. In the plane $z = z_{knot}$ magnetic field reaches its maximal value

$$\mathbf{H} = \mathbf{y}_0 \widetilde{H}_y(z_{knot}) e^{iqy - i\omega t},$$

where

$$\left|\widetilde{H}_y(z_{knot})\right|^2 = \left|\tilde{A}_0\right|^2 \kappa_\perp^2. \tag{45}$$

The field set by Eq. (44) has the following energy flux through the cross section of the waveguide:

$$\Pi = \frac{\omega}{8\pi} q l_\perp l_x \left|\tilde{A}_0\right|^2, \tag{46}$$

where $l_x$ is the waveguide size along the $x$ axis.

Using Eqs. (41) and (38) we obtain the following expression for the magnetic dipole losses at the resonant frequency $\omega = \omega_{21} + \Omega_{21}^{(A)}$ per unit area:

$$W^{(m)} \approx 2\alpha \cdot \frac{|z_{21}|^2 \omega^3}{c^2 \gamma} \cdot \frac{N_2^2 - N_1^2}{(N^*)^2} c \frac{\left|\widetilde{H}_y(z_{knot})\right|^2}{8\pi} \tag{47}$$

Summarizing Eqs. (45)-(47) we find the absorption rate of the considered waveguide mode per unit propagation length:

$$\mu = \frac{W^{(m)} l_x}{\Pi} \approx 2\alpha \cdot \frac{|z_{21}|^2 \omega^3}{c^2 \gamma} \cdot \frac{N_2^2 - N_1^2}{(N^*)^2} \frac{c\kappa_\perp^2}{\omega q l_\perp} \tag{48}$$

For the further estimations we choose $\bar{\varepsilon} \approx 10$ and $\frac{cq}{\omega} \approx 1$; in this case $\frac{c\kappa_\perp^2}{\omega q} \approx \bar{\varepsilon} - 1$. So, we can write the absorption coefficient for a waveguide section of the length $l_\parallel$ in the following form:

$$\mu l_\parallel \approx 4 \cdot 10^{-5} \left(\frac{|z_{21}|}{1\,nm}\right)^2 \left(\frac{\hbar\omega_{21}}{1\,eV}\right)^3 \left(\frac{10\,meV}{\hbar\gamma}\right) \frac{N_2^2 - N_1^2}{(N^*)^2} (\bar{\varepsilon} - 1) \frac{l_\parallel}{l_\perp}; \tag{49}$$

Thus, if we put a 5-layer QW of 3 mm lateral size in the waveguide of ~1.3 micron thickness (the wavelength of 0.3 eV photon in GaAs is 1.2 mkm), we get $l_\parallel/l_\perp \sim 4.6 \cdot 10^3$, and obtain 34% absorption of TE-mode. This estimation shows that the magnetic dipole absorption is measurable is a waveguide scheme. In Fig. 4 the theoretical absorption spectra of a quantum well for both the electric and magnetic dipole resonances are presented for the set of parameters listed above.

Note that optical cavities of similar geometry are widely used for fabrication of IR quantum cascade lasers [32].

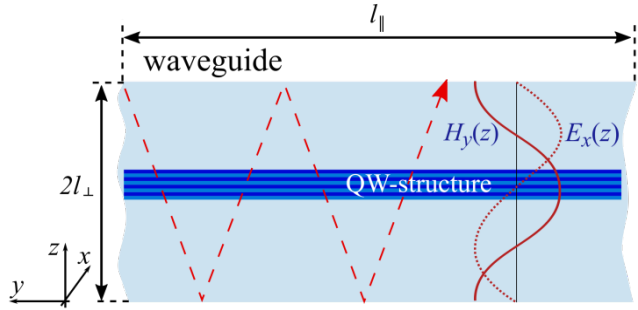

Fig 3. A possible scheme of the magnetic dipole absorption enhancement due to the multiple passing of light through QW in a planar waveguide; the case of $TE_{20}$-mode is shown, so the electric field (dotted red line) is parallel to the "$x$" axis. QW structure is placed in the plane of zero electric field, where the maximum of $y$-component of the magnetic field is localized (solid red line). Dashed red lines schematically show the trajectory of the light ray.

We should note that a waveguide providing "pure" TE mode is a theoretical idealization and even a weak component $\mathbf{E} \parallel \mathbf{z}_0$ will provide significant absorption. However, as we shown above, the relative frequency shift of magnetic and electric dipole resonances is greater than their widths, so these too effects can be measured separately. In Fig. 4 the absorption spectrum of the waveguide is shown for the case when the electric field component $E_z$ near the QW is equal to 0.5% of $H_y$.

Let us compare the magnetic dipole losses with the nonresonant Drude losses $W^{(Dr)}$. Using the relation for Drude absorption in QW shown in Fig. 3 under the condition $\gamma \ll \omega$ we find:

$$W^{(Dr)} \approx \frac{\gamma e^2 \sum_j N_j}{2 m_\parallel \omega^2} \langle |\tilde{E}_x|^2 \rangle, \tag{50}$$

where $\langle |\tilde{E}_x|^2 \rangle$ is the averaged squared electric field inside the QW:

$$\langle |\tilde{E}_x|^2 \rangle = \left(\frac{\omega}{c}\right)^2 \frac{1}{2L} \int_{z_{knot}-L}^{z_{knot}+L} |\tilde{A}_x(z)|^2 dz \approx \frac{L^2}{3} |\tilde{A}_0|^2 \left(\frac{\omega}{c}\right)^2 \kappa_\perp^2. \tag{51}$$

Taking into account Eq. (45), we obtain the following expression describing the electric field decrease near the cross-section $z = z_{knot}$:

$$\frac{\langle |\tilde{E}_x|^2 \rangle}{|\tilde{H}_y(z_{knot})|^2} = \frac{1}{3}\left(\frac{L\omega}{c}\right)^2 \tag{52}$$

Using Eqs. (47), (50) and (52), we get:

$$\frac{W^{(m)}}{W^{(Dr)}} \approx \frac{3|z_{21}|^2}{2L^2} \cdot \frac{\omega^2}{\gamma^2} \frac{N_2^2 - N_1^2}{N^* \sum_j N_j} \tag{53}$$

When $\frac{N_2^2 - N_1^2}{N^* \sum_j N_j} \sim 1$ and for the values $\frac{|z_{21}|}{L} \approx 0.34$, $\frac{\omega}{\gamma} \approx 15$ we find $\frac{W^{(m)}}{W^{(Dr)}} \approx 50$. It means that the magnetic dipole absorption is detectable against the background of Drude losses. Also note that the resonant nature of the magnetic dipole absorption makes it measurable against the background of even stronger non-resonant Drude absorption.

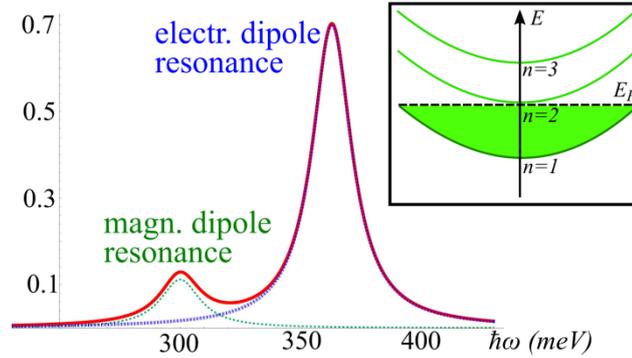

Fig. 4. The absorption spectrum of IR radiation in 5-layer QW placed inside the waveguide for the case of TE$_{20}$ mode with an admixture of TM component (electric field amplitude $E_z$ near QW is equal to 0.5% of the magnetic field amplitude $H_y$, the waveguide thickness is ~1.3 micron, the relative absorption per 1 mm of propagation distance is shown). Solid red line – total absorption, dashed green line – magnetic dipole absorption, dot-dashed blue line – electric dipole absorption. Basic parameters: GaAs ($m_\perp \approx m_\parallel = 0.067 m_0$, $\bar{\varepsilon} \approx 10$), photon energy $\hbar\omega \approx 0.3$ eV, $\hbar\gamma \approx 10\ meV$, the resonant transition between two lowest levels in a deep "rectangular" well is considered, QW thickness is 7 nm, $|z_{21}| \approx 1.2\ nm$. Inset: corresponding position of the Fermi level in QW (on the bottom of the second subband).

In general case, subband energies $E_m$ used above should be calculated taking into account the deformation of the potential well $U_{qw}(z)$ due to the redistribution of charge carriers and the exchange effects (for example, it can be described in the Hartree-Fock approximation [11, 33]). However, the exchange effects modify the frequencies of the electric dipole resonances $\omega = \frac{E_m - E_n}{\hbar} + \Omega_{mn}^{(\varphi)}$ not only through the intersubbang transition frequencies $\frac{E_m - E_n}{\hbar}$, but also trough the influence on "plasma" frequency shifts $\Omega_{mn}^{(\varphi)}$.

At the same time, corresponding frequency shifts of the magnetic dipole resonances $\Omega_{mn}^{(A)}$ do not depend on the exchange effects since they do not depend on Coulomb interaction at all. As we showed above, for typical QW parameters the frequencies of the magnetic dipole resonances almost do not shift relative to the intersubband transition frequencies, while the electric dipole resonances are significantly modified due to the plasma and exchange effects. Thus, the above estimations show that the magnetic dipole mechanism of absorption is a relatively weak effect, but probably it can be observed experimentally. As we mentioned previously, measurement of the frequency difference between the magnetic and electric dipole resonances may be used both for the QW parameters diagnostics and the investigation of collective effects.

**Acknowledgements**

The authors thank A. Belyanin and I. D. Tokman for helpful discussions. The work was supported by the Russian Foundation for Basic Research (RFBR, Project No. 18-29-19091 mk); also I.V.O. is grateful to the Foundation for the Advancement of Theoretical Physics and Mathematics "BASIS" for support.

## APPENDIX
**Influence of the spin effects on the absorption of normally incident radiation.**

Taking into account spin effects leads to the modification of the perturbation operator. First of all, the energy of the spin magnetic moment in a magnetic field should be included: $\hat{V}_B = -2\mu_B(\hat{\boldsymbol{s}} \cdot \boldsymbol{B})$, where $\mu_B$ is the Bohr magneton, $\hat{\boldsymbol{s}} = \frac{1}{2}(\boldsymbol{x}_0\hat{\sigma}_x + \boldsymbol{y}_0\hat{\sigma}_y + \boldsymbol{z}_0\hat{\sigma}_z)$ is the spin operator. In the case under consideration the direction of the magnetic field $\boldsymbol{B} = \nabla \times \boldsymbol{A}$ is time independent, so the operator of the spin projection on the quantization axis commutes with the Hamiltonian $\hat{H}$ (see [15]).

Further, let us consider the spin-orbital interaction. We will use a well-known expression for the corresponding interaction operator [16, 34]: $\hat{V}^{(s-o)} = \frac{e\hbar}{2m_0^2 c^2}[\boldsymbol{E} \times \hat{\boldsymbol{p}}] \cdot \hat{\boldsymbol{s}}$. For the electric field $\boldsymbol{E} = \boldsymbol{x}_0 \mathrm{Re}\tilde{E}_x e^{-i\omega t}$ we obtain:

$$\hat{V}^{(s-o)} = \frac{e\hbar^2 \tilde{E}_x}{4m_0^2 c^2}\left(-i\hat{s}_y \frac{\partial}{\partial z} + \hat{s}_z k_y\right) e^{-i\omega t} + H.C. \tag{A1}$$

The simplest estimation of the absorption due to this effect is based on the Einstein coefficient method[3] [35]. Probability of the photon absorption and the induced emission can be expressed through the spontaneous emission probability. To find the latter use the Fermi golden rule [1, 15]; this probability should be proportional to $\left|V^{(s-o)}_{mnss'kk}\right|^2$ (where $s, s'$ are the spin state indices). Defining the semi-width of the resonant frequency band as $\gamma$, we get the following power loss per unit area:

$$W^{(s-o)} \approx \frac{\omega}{2} \frac{\sum_{kss'}\left|V^{(s-o)}_{mnss'kk}\right|^2 \left(n^{(0)}_{nsk} - n^{(0)}_{ms'k}\right)}{\hbar \gamma}. \tag{A2}$$

Using Eq. (A2) and Eq. (41) for the power loss at the magnetic dipole resonance we find:

$$\frac{W^{(s-o)}}{W^{(m)}} \approx \frac{m_\parallel}{16\varepsilon_{(-)}m_0} \cdot \frac{1}{m_0 c^2}\left(\frac{m_\perp^2 \hbar^2 \omega_{mn}^2}{m_0^2 W^{mn}_{eff}} + \frac{\hbar^2}{m_0 |z_{mn}|^2}\right). \tag{A3}$$

From Eq. (A3) it follows that the absorption due to spin-orbital interaction is comparable with the magnetic dipole absorption just in the case of near-zero carrier density. At the same time, in narrow-gap materials the spin-orbital interaction due to the action of crystal lattice field is an efficient mechanism of multiband coupling [36].

---
[3] Of course, in this way we could also obtain Eq. (41)